\documentclass[journal=ancac3,manuscript=article]{achemso} 

\usepackage{chemformula} 
\usepackage[T1]{fontenc}

\usepackage{graphicx}
\usepackage{dcolumn}
\usepackage{bm}
\usepackage{hyperref}
\usepackage {xcolor} 
\usepackage{amsmath,amssymb}

\usepackage[caption = false]{subfig}
\usepackage{epstopdf}

\author{Bernat Dur{\`a} Faul{\'i}}
\affiliation{Secci\'o de F\'isica Estad\'istica i Interdisciplin\`aria - Departament de F\'isica de la Mat\`eria Condensada, Universitat de Barcelona, Mart\'{\i} i Franqu\`es 1, 08028 Barcelona, Spain.}
\author{Valentino Bianco}
\affiliation{Onena Medicines S.L.,
Paseo Miram\'on, 170, planta 3, B06, 20014, Donostia, Gipuzkoa,
Spain.}
\author{Giancarlo Franzese}
\affiliation{Secci\'o de F\'isica Estad\'istica i Interdisciplin\`aria - Departament de F\'isica de la Mat\`eria Condensada, Universitat de Barcelona, Mart\'{\i} i Franqu\`es 1, 08028 Barcelona, Spain.}
\alsoaffiliation{Institut de Nanoci\`{e}ncia i Nanotecnologia, Universitat de Barcelona, 08028 Barcelona, Spain.}
\email{gfranzese@ub.edu}

\title[]
  {Hydrophobic homopolymer's coil-globule transition and adsorption onto a hydrophobic surface under different conditions}


\begin{document}

\date{\today}

\begin{abstract}
Unstructured proteins can modulate cellular responses to environmental conditions by undergoing coil-globule transitions and phase separation. However, the molecular mechanisms of these phenomena still need to be fully understood.
Here, we use Monte Carlo calculations of a coarse-grained model incorporating water's effects on the system's free energy.
Following previous studies, we model an unstructured protein as a polymer chain. Because we are interested in investigating how it responds to thermodynamic changes near a hydrophobic surface under different conditions, we choose an entirely hydrophobic sequence to maximize the interaction with the interface.
We show that a slit pore confinement without top-down symmetry enhances the unfolding and adsorption of the chain in both random coil and globular states.
Moreover, we demonstrate that the hydration water modulates this behavior depending on the thermodynamic parameters.
Our findings provide insights into how homopolymers, and possibly unstructured proteins, can sense and adjust to external stimuli such as nanointerfaces or stresses.
\end{abstract}

\maketitle

\section{Introduction}

Structured proteins fold into a specific three-dimensional structure to achieve their function. However, proteins with intrinsically disordered regions (IDRs) and intrinsically disordered proteins (IDPs) have regions or domains that remain unfolded or disordered under physiological conditions. IDRs larger than 30 amino acids and IDPs are common in cells and regulate diverse cellular processes, such as RNA binding, oligomerization, metabolite recruitment, and catalysis \cite{Zhou:2018vt}.
Moreover, IDRs and IDPs are exposed to weak, multivalent, and dynamic interactions that could lead to liquid-liquid phase separation (LLPS), a phenomenon in which they form droplet-like structures that concentrate biomolecules without a membrane barrier. The biomolecular condensation potentially involves various biological functions and dysfunctions, such as gene regulation, signal transduction, and neurodegeneration \cite{Alberti:2021vv, biom12091266}. 
Interestingly, IDRs and IDPs can phase-separate at much lower concentrations than structured proteins, such as those involved in cataract formation or fibrils. 
However, the balance between liquid-like and solid-like phases is delicate and depends on the type of interaction among the disordered molecules. For example, homotypic interactions tend to promote aggregation and fibrillation, which can be detrimental to cellular health. On the other hand, heterotypic interactions can stabilize the liquid phase and prevent pathological phase transitions \cite{Mathieu56}.

Recent studies have linked IDPs' coil-globule transition to their LLPS as a function of the protein sequence. This allows the calculation of sequence-specific phase diagrams \cite{Zeng:2020aa}. 
Another elegant work, coarse-graining multiple IDP amino acids as beads on a string, discovered a surprisingly rich phase separation behavior by changing the sequence \cite{Statt:2020vu}.  For sequences mainly hydrophobic, the authors found conventional LLPS and a re-entrant phase behavior for sequences with lower hydrophobicity.
It is, therefore, interesting to explore how heterotypic interactions of IDPs can affect their sequence-dependent coil-globule transition (and condensation) using simple models.

Furthermore, in many fields like medicine \cite{10.3389/fbioe.2020.00990, https://doi.org/10.1002/adma.201804922,D0CS00461H, Bhatia2022}, food science \cite{nano9020296, foods9020148, MORADI202275}, and biosensors \cite{LAN2017504, Zhu2015, C9NA00491B}, it is essential to understand how proteins and biomolecules interact with nanomaterials. 
For example, when nanoparticles come into contact with bloodstreams, they form a corona of multiple layers of proteins and biomolecules. This gives the nanocomplex a new biological identity \cite{https://doi.org/10.1002/wnan.1615, Vilanova2016}.
It is generally accepted that, upon adsorption, proteins can alter their structure  \cite{Fernndez2001OnAD, Rabe:2011kx, Wheeler2021}, which can have significant consequences like an inflammatory response or fibril formation \cite{Park:2021ve, Li:2019aa}.
However, our comprehension of these mechanisms must still be completed \cite {Assfalg:2021aa}.
Also, the effect of adsorption on a flat surface can be highly diverse when comparing structured regions with IDRs of the same protein \cite{Desroches:2007aa}.
Hence, understanding the impact of the interface on the protein's conformation is crucial in determining nanomaterial interactions with biological environments \cite{Vilanova2016, C1CS15233E, FranzeseBiancoFoodBio2013, polym13010156, CHEN2022114336, D1NR06580G}.

Here we consider the coarse-grained Bianco-Franzese (BF) model for proteins in explicit water in its simplest version \cite{Bianco:2015aa, FranzeseBiancoFoodBio2013}, as defined below. Despite its schematic approximations, the BF model can show, both for structured proteins with a native state and for IDPs, that accounting for the contribution of the hydration water \cite{Franzese2011}
is enough to predict protein thermodynamic properties
consistent with theories \cite{hawley1971, Bianco2012a}
and experiments \cite{Granata:2015aa, Bianco:2017aa}.

The BF Hamiltonian model reproduces, for both structured and unstructured proteins, elliptically-shaped stability regions (SRs) in the temperature-pressure ($T$-$P$) plane \cite{Gross1994, Lesch2002}.
The SRs include high-$T$ unfolding (melting), driven by the entropy increase, which is common to all the protein models, e.g., Ref. \cite{Lau1989}.
Additionally, the BF model shows that hydration-water energy drives the low-$T$ (cold) unfolding. Hydrophobic-collapse models cannot explain this experimental phenomenon \cite{PhysRevE.61.R2208, PhysRevE.63.031901}. 
Specific models can reproduce the cold unfolding without \cite{Dias2008, Yoshidome} or with \cite{Marques2003, Patel2007} a $P$-dependent behavior. However, at variance with the BF model, they do not reproduce the experimental elliptic SR.

Moreover, the BF model explains high-$P$ unfolding as density-driven due to increased hydration water compressibility at hydrophobic interfaces \cite{Dadarlat2006, Godawat2009, Sarupria:2009ly, Das2012}, common also to other water-like models \cite{Strekalova2012}.
Finally, it explicates the low-$P$ denaturation seen in the experiments \cite{Lesch2002, Larios:2010aa} and models \cite{Hatch:2014aa}
as enthalpy-driven \cite{Bianco:2015aa}.

The BF model has other interesting properties. For example, it sheds light on water's evolutionary action in selecting protein sequences and the effect of extreme thermodynamic conditions. This has implications for protein and drug design \cite{BiancoPRX2017}. 
For example, the model shows that artificial covalent bridges between amino acids are necessary to avoid protein denaturation at $P>0.6$ GPa \cite{Lesch2002}. 
Moreover, it also helps us understand why only about 70\% of the surface of mesophilic proteins is hydrophilic, and about 50\% of their core is hydrophobic \cite{BiancoPRX2017}.

Recently, the BF model has been used to study how structured proteins denature and aggregate reversibly depending on their concentration in water solutions with one \cite{Bianco:2020aa} or two protein components \cite{Bianco-Navarro2019} or near hydrophobic interfaces \cite{polym13010156}. 
The results show that unfolding facilitates reversible aggregation  \cite{Bianco:2020aa} with a cross-dependence in multicomponent mixtures \cite{Bianco-Navarro2019}. Also, the proteins aggregate less near hydrophobic interfaces, at high $T$, or by increasing the hydrophobic effect (e.g., by reducing salt concentration) \cite{polym13010156}.

Hydrophobic slit-pore confinement has been extensively studied for polymers near the coil-globule transition adopting lattice models. For example, it has been disputed if the collapse temperature has a maximum at a specific slit-pore inter-wall separation \cite{Mishra:2004aa} or if it just increases monotonically \cite{Hsu_2005}, with recent results \cite{Dai:2015aa} possibly reconciling the debate based on the ratio between the chain length and the slit-pore size.

Here, we study by Monte Carlo calculations on a compressible lattice model in two dimensions (2D) how adsorption on a hydrophobic wall (a line in 2D) of a slit-pore affects the coil-globule transition of an unstructured, entirely hydrophobic homopolymer, used here as the simplest model for a hydrophobic protein. Our slit pore has a fixed size, which is larger than the maximum extension of the protein. Therefore, the polymer can interact only with one wall at a time, allowing us to assume that the farthest wall is not reducing the number of visited configurations, as it would be if a protein were near a single interface. 
The results help us to understand the fate near nanomaterials of hydrophobic homopolymers and, possibly, unstructured proteins \cite{trp, china}.

\section{Model}

\subsection{The FS model for water}

The BF model is based on adding a coarse-grained protein with its hydrated interface to the Franzese-Stanley (FS) water model \cite{Franzese:2002aa, FS-PhysA2002, FMS2003, FS2007}.
The FS model includes cooperative (many-body) interactions in an effective lattice-cell model proposed by Satsry et al. \cite{Sastry:1996aa} with only two free parameters: 1) $J/\epsilon$ quantifying the relative strength of the directional component of the hydrogen bond (HB) interaction $J$ compared to van der Waals interaction parameter $\epsilon$,
and 2) the HB-dependent cell-volume variation $v_{\rm HB}$ expressed in units of  the water van der Waals volume $v_0$. 
The FS model adds a third parameter, $J_\sigma/\epsilon$, describing the HBs cooperativity and indicating the strength of many-body HBs in van der Waals units. The ratio $J_\sigma/J$ controls the phase diagram in the supercooled region \cite{Stokely2010}. 

The FS model coarse grains the water atomistic coordinates, introducing a density field with local fluctuations due to the HB structure but keeping a molecular description of the HB network. Recent reviews summarize the definition of the FS model for a water monolayer and its main properties \cite{Gallo:2021wx, CoronasBookMartelli2022}.

The extension of the FS model to bulk shows that its three parameters can be adjusted in a way to give optimal agreement with the experimental water data in an extensive range of $T$ and $P$ around ambient conditions \cite{LuisThesis2023} (for preliminary calculations, see Ref. \cite{Coronas2016}). 
However, the HB network's peculiar structure that preferentially has a low (four) coordination number makes the 2D monolayer version of the model, with only four neighbors, interesting.
Indeed, the FS 2D monolayer offers a reasonable coarse-grained approximation for the water equation of state near ambient conditions at the cost of renormalizing its parameters. This renormalization allows us to account for the difference in entropy compared to the bulk, with the advantage of being easier to visualize and calculate. 

Therefore, we consider a partition of the system's 2D-projection into $N$ square cells, of which water molecules occupy $N_W\leq N$, 
each with the average volume $v(T, P) \geq v_0$, the van der Waals excluded volume for a water molecule without HBs.
On the other hand, we assume that the
HBs are the primary source of local density fluctuations and associate with each HB a proper volume $v_{\rm HB}/v_0=0.5$ equal to the average
volume increase per HB between high-density ices VI and VIII and
low-density (tetrahedral) ice Ih, approximating the average volume variation per HB when a tetrahedral HB network is formed \cite{Soper-Ricci-2000}. 
Hence, the volume of water is
\begin{equation}
V\equiv N_Wv+N_{\rm HB}v_{\rm HB},
\label{Vol}
\end{equation}
where $N_{\rm HB}$ is the number of HBs.

The FS Hamiltonian, describing the interaction between the water molecules,  is
\begin{equation}
 {\cal H}_{W,W} \equiv \sum_{ij} U \left( r_{ij} \right) - JN_{\rm HB} - J_{\sigma}N_{\sigma},
\label{H_WW}
\end{equation}
where 
$U=\infty$ 
for 
$r<r_0\equiv v_0^{1/3}=2.9$ \AA, and 
\begin{equation}
U(r)\equiv 4\epsilon \left[ \left(\frac{r_0}{r}\right)^{12} - \left(\frac{r_0}{r}\right)^{6}\right]  
\label{U}
\end{equation}
for 
$r_0 < r < 6r_0$ (cutoff) or 
$U=0$ for larger $r$, 
with $\epsilon =5.8$ kJ/mol. 
The sum runs over all possible water-molecule couples (including those in the hydration shell introduced in the BF model) and is not limited to nearest neighbor (NN) molecules.
This term
accounts for the O--O van der Waals interaction between molecules at a distance $r$. 
It differs from the squared-well interaction used in the original formulation of the Sastry et al. model \cite{Sastry:1996aa} and in the mean-field solution of the FS model \cite{FS2007}.
This difference allows for the continuous change of the distance between the cell centers and their volume, making the lattice compressible and the model suitable for calculations at different pressures $P$. Together with the local volume fluctuations allowed by the Eq.~(\ref{Vol}), the continuous choice for $U$ allows for better matching of the model's equation of state to the water case, curing the lattice artifacts related to the flatness of the bottom of the potential curve and the fixed lattice spacing.  
Previous calculations prove that the model results are independent of the details of $U$ \cite{Santos:2011aa}.

Because we consider the system at constant $NPT$, the distance $r_{ij}$ is a continuous variable. 
Notably, because the formation of HBs does not change the  NN distance
$r_{ij}^{(NN)}/r_0\equiv (v/v_0)^{1/3}$
between water molecules in the first coordination shell 
\cite{Soper-Ricci-2000}, the van der Waals interaction is unaffected by the HBs, guaranteeing that the FS is not just a simplified mean-field model.

The term $-JN_{\rm HB}$ accounts for the additive (two-body) component of the HB. 
The FS model adopts the HB definition based on the distance between the centers of mass of two water molecules and the angle between the OH group of one and the O atom of the other \cite{Franzese:2002aa}
The HB has minimum energy when the H is along the O-O direction or deviates less than $30^o$ \cite{CoronasBookMartelli2022, Ferrario:1990aa, Luzar-Chandler96}.
Hence, only 1/6 of all the possible orientations in the plane of the H atom relative to the O-O direction corresponds to a bonded state, while the other 5/6 states are non-bonded. 
Therefore, to correctly account for the entropy variation once the HB is formed, we introduce a 6-state bonding variable 
$\sigma_{ij}$ for each of the four possible HBs that each water molecule $i$ can form with a NN water molecule $j$. 
We assume that the HB is formed only if both molecules have the same bonding state, i.e., if $\delta_{\sigma_{ij}, \sigma_{ji}}=1$, where $\delta_{ab}=1$ if $a=b$, 0 otherwise.

On the other hand, the HB can be considered broken when the O-O is larger than a given $r_{\rm max}$ \cite{Hus:2012uq}. 
The FS model assumes the reasonable value $r_{\rm max}\simeq 3.65$~\AA \cite{CoronasBookMartelli2022}, implying that for $r>r_{\rm max}$ it is $(r_0/r)^3\equiv v_0/v < 0.5$. 
Hence, by setting $n_i=n\equiv \theta(v_0/v - 0.5)$, where $\theta(x)$ is the Heaviside step function, the total number of HBs is $N_{\rm HB} \equiv \sum_{<i,j>}n_in_j\delta_{\sigma_{ij},\sigma_{ji}}=
\theta(v_0/v - 0.5)\sum_{<i,j>}\delta_{\sigma_{ij},\sigma_{ji}}$.

The last term in the Hamiltonian, $-J_{\sigma}N_{\sigma}$, accounts for the many-body term that can be calculated by {\it ab initio} methods. It favors the formation of a low-density (tetrahedral) local structure in liquid water even at ambient conditions \cite{Skarmoutsos:2022ur}.
In classical atomistic potentials,
this term is modeled with a long-range polarizable dipolar interaction. However, recent calculations, based on polarizable models including the MB-pol potential \cite{Babin:2013vt, Babin:2014vf, Medders:2014wb,bore_paesani_2023}, show that it can be approximated with a short-range 5-body interaction within the first coordination shell of a water molecule \cite{Abella:2023aa}.
This result gives a solid theoretical foundation to the FS assumption of modeling the cooperative term as an effective 5-body interaction within the first coordination shell of each water molecule $i$, with 
$N_\sigma \equiv \sum_i \sum_{(k,l)_i}\delta_{\sigma_{ik},\sigma_{il}}$, where the inner sum is over all the pairs of the bonding variables of the molecule $i$.
Following \cite{Bianco:2015aa}, we set here $J/4\epsilon=0.3$ and
$J_\sigma/4\epsilon = 0.05$.

\begin{figure}
\centering
\includegraphics[width=\columnwidth]{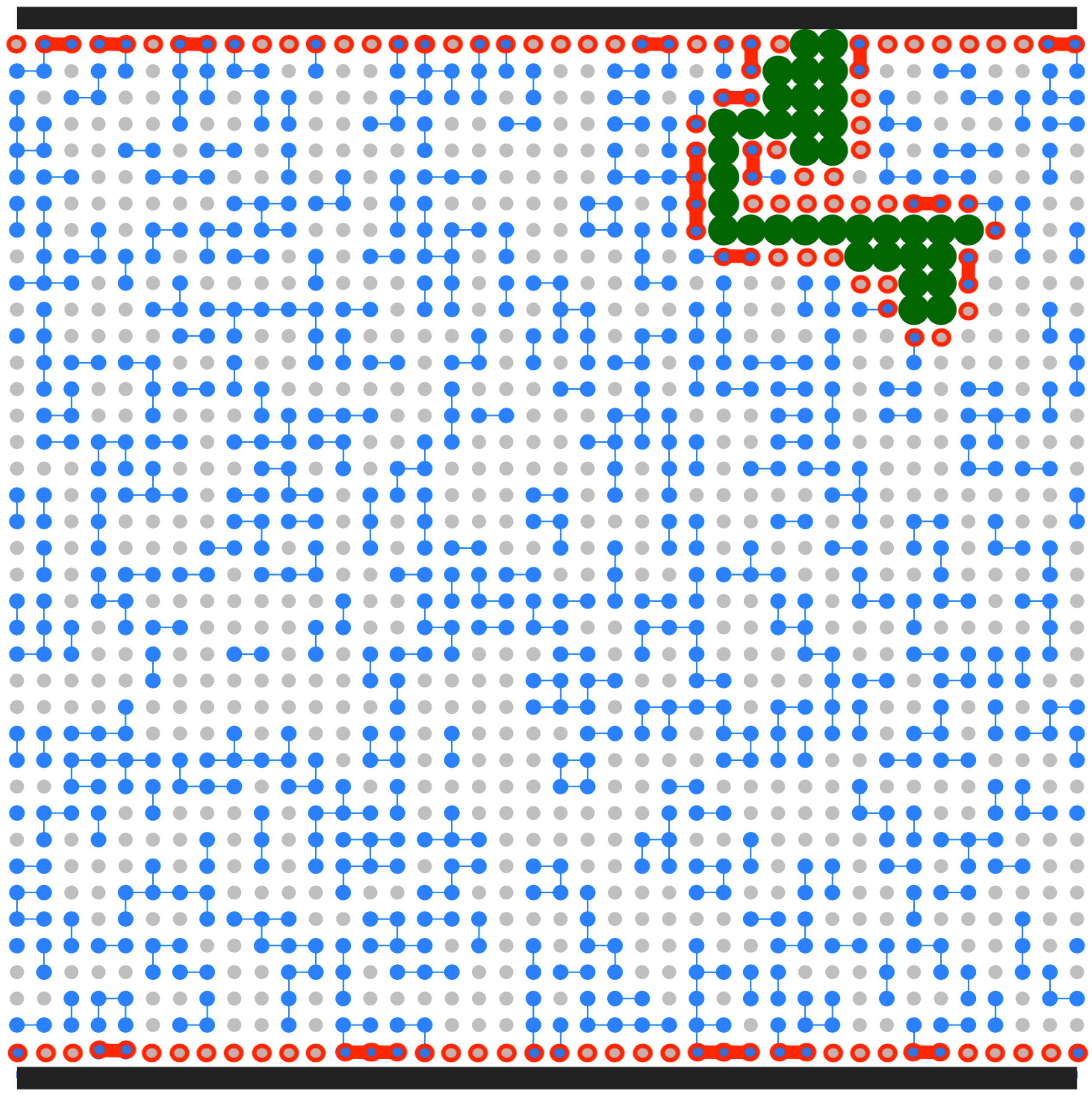}
\caption{{\bf Example of one of the visited conformations for a homopolymer confined in a slit-pore.} The homopolymer (chain of green beads), coarse-graining an unstructured protein backbone in the BF model, is adsorbed onto one of the pore's walls (black top and bottom lines) and surrounded by water (dots). At the thermodynamic conditions of this example
($k_BT/4\epsilon$=0.55 and $Pv_0/4\epsilon$=0.4),
some water molecules  (grey dots) do not form any HB, while others (blue dots) can have up to four HBs (blue lines).  Water molecules and water-water HBs in the (protein and wall) hydration shells are highlighted in red.}
\label{system}
\end{figure}

\subsection{The BF model for a hydrophobic homopolymer}

Based on atomistic results, the BF model assumes that in a hydrophobic ($\phi$) hydration layer (Fig. \ref{system})
\begin{itemize}
\item[{\bf i)}] 
the interfacial water-water HBs are stronger than bulk HBs, with an extra interaction $\Delta J^{(\phi)}/J=0.83$;
\item[{\bf ii)}] the water compressibility is larger than bulk compressibility \cite{Dadarlat2006, Godawat2009, Sarupria:2009ly, Das2012}, so that HB's volume is reduced by 
$\Delta v_{\rm HB}^{(\phi)}/v_{\rm HB} = - k_1P$, with $k_1=v_0/4\epsilon$.
\end{itemize}

Hence,  the FS enthalpy $H_{W,W}^{(FS)}\equiv {\cal H}_{W,W} +P V$, from Eq.s (\ref{Vol},   
 \ref{H_WW}), acquires an extra term in the BF model that  
for the hydrophobic hydration shell is 
\begin{equation}
\Delta H^{(h)}_{W,W} = - (\Delta J^{(\phi)} +k_1P^2v_{\rm HB})N_{\rm HB}^{(\phi)},
\label{BF_water}    
\end{equation}
where $N_{\rm HB}^{(\phi)}$ is the number of HBs between water molecules in the hydration shell. 

Although we have implemented a high-resolution version of the BF model \cite{LuisThesis2023}, here we adopt a simple coarse-grain representation of  beads-on-a-chain, with one bead per residue, 
that has been extensively used in the literature to get a qualitative understanding of protein properties \cite{BiancoPRX2017}. 
The protein-like polipetyde Hamiltonian
\begin{equation}
{\cal H}_p \equiv {\cal H}_{R,R} + {\cal H}_{R,W}
\label{prot}
\end{equation}
describes the interactions among the NN residues, ${\cal H}_{R, R}$, and between the residues and the NN water molecules in the hydration shell, ${\cal H}_{R, W}$ \cite{BiancoPRX2017}.
Here we represent an unstructured protein with a hydrophobic homopolymer where all the $N_R$ residues interact with the NN molecules by excluded volume. A more general expression for ${\cal H}_p$ accounting for the complete protein amino acids is presented in Ref.s~\cite{Bianco:2015aa, Bianco:2017aa, BiancoPRX2017}.
The model parameters are chosen in such a way as to mimic pH and salt conditions at which there are no long-ranged electrostatic interactions, and the Hamiltonian only has short-range terms. 

Finally, the BF enthalpy of the entire system with the hydrated protein in explicit water is
\begin{equation}
H^{(BF)} \equiv H_{W,W}^{(FS)} +\Delta H^{(h)}_{W,W} + {\cal H}_p.
\label{BF}
\end{equation}

The general expression for the Gibbs free energy of the BF model is
\begin{equation}
G^{(BF)}\equiv {\cal H}_{\rm TOT} + P V_{\rm TOT} - TS_{\rm TOT},
\label{GBF}
\end{equation}
where 
${\cal H}_{\rm TOT} \equiv {\cal H}_p + {\cal H}_{W,W} 
- \Delta J^{(\phi)} N_{\rm HB}^{(\phi)}$,
$V_{\rm TOT}\equiv V-k_1Pv_{\rm HB}N_{\rm HB}^{(\phi)}$,
and $S_{\rm TOT}$ is the total entropy of the system associated with all the configurations having the same number of proteins contact points (CPs), $N_{\rm CP}$, defined in the following, and the same number of water molecules in the hydration shell (red dots in Fig.~\ref{system}).

As in the BF original formulation, we assume that 
protein residues and water molecules have the same size. Recently, we have developed a version of the model in which we remove this limitation by letting each residue occupy several cells, where the cells have the size of a water molecule \cite{LuisThesis2023}. This modification leads to a high-resolution lattice model with conformation indeterminacy comparable to coarse-grained (CG) water-implicit models \cite{C4CS00048J}.
Regarding the free-energy calculations in bulk, we find \cite{LuisThesis2023} that the main effect of the high-resolution lattice is to increase the hydrated protein surface. This observation implies that, by rescaling the model's parameters for the hydration energy and entropy, there is no qualitative change in the free-energy calculations in bulk. 
On the other hand, entropic effects could be different near a surface due to the limitation of accessible conformations. 
However, the reduced number of acceptable water configurations in 2D should reduce the entropy difference between the low and high-resolution cases, preserving the qualitative agreement we seek in this work. This argument is supported by our results being qualitatively consistent with those for confined polymers. 
Further studies beyond the scope of the present work are needed to answer this question in more detail. 

\subsection{Monte Carlo calculations with and without top-down symmetry}

We realize the slit pore geometry in a square partition with size $L=40$
by fixing $L$ hydrophobic cells along a line and applying periodic boundary conditions in all directions
(Fig. \ref{system}). 
We perform Monte Carlo (MC) calculations for a protein-like chain with $N_R=36$
residues at constant $P$, $T$, $N_W$ and $N_R$, with $N_W+N_R=N\equiv L^2$.

We consider random initial configurations and equilibrate the water bonding indexes with a clustering algorithm \cite{Mazza2009} and the chain with corner flips, pivots, crankshaft moves, and random unitary translations of its center of mass \cite{Understanding-Frenkel}. 
A single MC step is made of a random sequence of move-attempts for each degree of freedom of the system (36 residues and 6256 $\sigma_{ij}$ variables).
After moving the chain, the cells left by the amino acids are replaced by water molecules whose values of the four $\sigma_{ij}$ variables are chosen randomly  \cite{Bianco2012a, Vilanova:2017aa, polym13010156}. 

To facilitate the protein-like polymer adsorption, we break the top-down symmetry by biasing the translation toward one of the confining walls but not along the slit pore. For the sake of the description, we call {\it top} the biased wall. 
The bias mimics a drift or a weak force pushing a protein toward the top interface without limiting its thermal motion parallel to the walls.
In the Supplementary Material, we discuss the case without bias, i.e., with top-down symmetry.

We perform calculations for temperatures ranging from 
$k_BT/4\epsilon$=0.01 to 0.6 and pressures from $Pv_0/4\epsilon=-0.2$ to 0.6. 
For each $(T, P)$, we collect configurations for every 100 of $10^6$ MC steps after discarding $10^4$ equilibration steps. 

Our main observable is how close the chain is to a globule conformation. 
To this goal, we calculate the degree of folding as the number $N_{\rm CP}$ of contact points (CP) that the polymer has with itself. 
We consider that there is a CP if two residues occupy the NN cells but are not adjacent along the chain. 

We remark that our MC polymer configurations are generated on a square lattice because lattice MC models can sample accessible conformations much more efficiently than the off-lattice counterparts. This is due, e.g., to the CG representation of the chain, a discretized number of bond vectors, and a higher fraction of acceptances of MC moves through easy identification of overlaps \cite{Gartner:2019aa}. Therefore, lattice models allow studying problems at considerable length and time scales where atomistic or off-lattice CG models are not feasible. 

However, the lattice dictates the distribution of bond lengths and angles, affecting the accessible conformations in an artificial way \cite{Gartner:2019aa}.
Furthermore, such a model captures only the configurational part of the partition function and does not allow one to calculate the forces and momentum, particularly if it has the bottom of the potential curve flat and the width adjusted to the lattice spacing \cite{https://doi.org/10.1002/jcc.540141009}. 
Importantly, these limitations apply only partially to our model that, instead, has a continuous interaction potential because the lattice cells are compressible and the distance between the monomers changes continuously, as in Eq.~(\ref{U}). This feature allows us also to perform constant pressure simulations, an option not available in incompressible lattice models \cite{Gartner:2019aa}.

Despite lattice anisotropy artifacts could be severe, it has been observed excellent agreement between the off-lattice and lattice results for many measured quantities, including the gyration radii \cite{Halverson_2013}.
On-lattice self-avoiding random walks provide a good approximation for the coil-globule transition and capture some essential features of the all-or-none folding transition of small globular proteins \cite{Kolinski2011}.
For example, Levitt adopted a $6\times6$ 2D-square lattice model to construct test proteins and extract knowledge-based energy functions for them \cite{10.1063/1.1320823}, Buchler and  Goldstein \cite{10.1063/1.480893} and Li et al. \cite{https://doi.org/10.1002/prot.10239} used 2D-square lattice models of similar size
to investigate questions about protein structure designability, up to more recent applications of lattice models for simulating phase transitions of multivalent proteins by Pappu and coworkers \cite{Choi:2019wy}.
Furthermore, the resolution of lattice models can vary from a very crude shape of the main chain to a resolution similar to that of good experimental structures \cite{Kolinski:2004aa, LuisThesis2023}.
Low-resolution lattices, in which the sites connected by site-site virtual bonds are located on NN lattice nodes, can be used only to study protein-like polymers. In contrast, in high-resolution lattices, the accuracy can be as high as 0.35\AA, comparable to that of CG force fields\cite{Kolinski2011}.

Although we have implemented such high resolutions in our approach \cite{LuisThesis2023}, here we adopt a low-resolution lattice model of protein-like polypeptides, following the hydrophobic and polar (HP) model proposed by Dill and coworkers \cite{Lau1989} and extensively studied, e.g., in Ref.s 
\cite{doi:10.1073/pnas.87.16.6388, 
10.1063/1.466677, 
https://doi.org/10.1002/pro.5560040401, 
doi:10.1073/pnas.92.1.146, 
SALI19941614, 
Sali:1994aa,
doi:10.1073/pnas.93.16.8356, 
doi:10.1146/annurev.biophys.30.1.361}.
This choice is computationally very efficient and allows us to qualitatively analyze the different contributions of the system's free energy.

\begin{figure}
\centering
\includegraphics[width=\columnwidth]{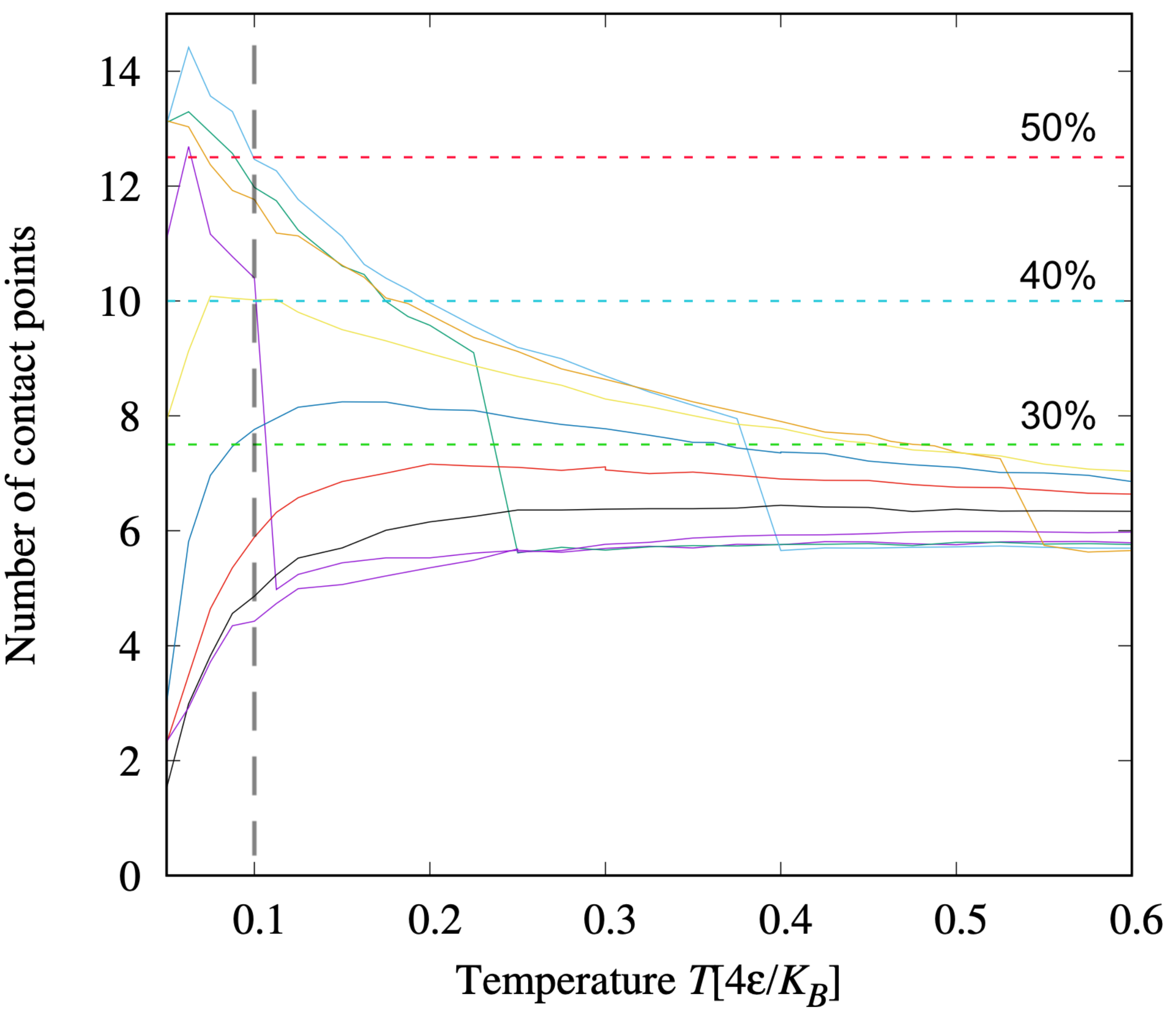}
\caption{{\bf Coil-globule transition for the hydrophobic homopolymer in the hydrophobic slit pore without top-down symmetry.}
The number of CPs, $N_{\rm CP}$, at constant $P$ as a function of $T$ is non-monotonic for any $Pv_0/4\epsilon<0.5$, showing a reentrant coil-globule transition when we consider CP's thresholds at 50\%, 40\%, and 30\% of CP$_{\rm max}$ (red, blue, and green horizontal dashed lines, respectively).
The calculations are presented as segmented lines (points connected by segments) for 
pressures, from bottom to top at $k_BT/4\epsilon =0.1$ (vertical dashed grey line), $Pv_0/4\epsilon=$ 
0.6 (indigo),
0.5 (black),
0.4 (red),
0.3 (blue),
0.2 (yellow),
-0.2 (indigo),
0.1 (orange),
-0.1 (green),
0.0 (turquoise). 
Note that the negative pressures intercalate with the positive.
Discontinuities for the four lowest pressures mark the limit of the liquid-to-gas spinodal of the confined water solution when $T$ increases.}
\label{CP}
\end{figure}

\section{Results and discusion}

\subsection{Coil-globule transition}

For each $(T, P)$, we compute the average $N_{\rm CP}$ for the chain.
For our 36 residue-long polypeptides, the maximum $N_{\rm CP}$ is $N_{\rm CP}^{\rm max} = 25$. 
When $N_{\rm CP}> 50\% N_{\rm CP}^{\rm max}$, we identify the conformation as globule, while we consider it coil otherwise (Fig.\ref{CP}).
Our calculations show that, for $Pv_0/4\epsilon<0.5$, $N_{\rm CP}$ is non-monotonic as a function of $T$. 
For $Pv_0/4\epsilon= 0.3$ (blue line in Fig.\ref{CP}), $N_{\rm CP}$ is larger than $30\% N_{\rm CP}^{\rm max}$ in a limited range of $T$, but does not reach the 40\% threshold. 
Within our resolution of $P$, $Pv_0/4\epsilon= 0.2$ (yellow line in Fig.\ref{CP}) is the highest at which the chain reaches $40\% N_{\rm CP}^{\rm max}$, while for any $Pv_0/4\epsilon\leq 0.1$ (orange line in Fig.\ref{CP})
 it undergoes a coil-globule transition.
 
 \begin{figure}
\centering
\includegraphics[width=\columnwidth]{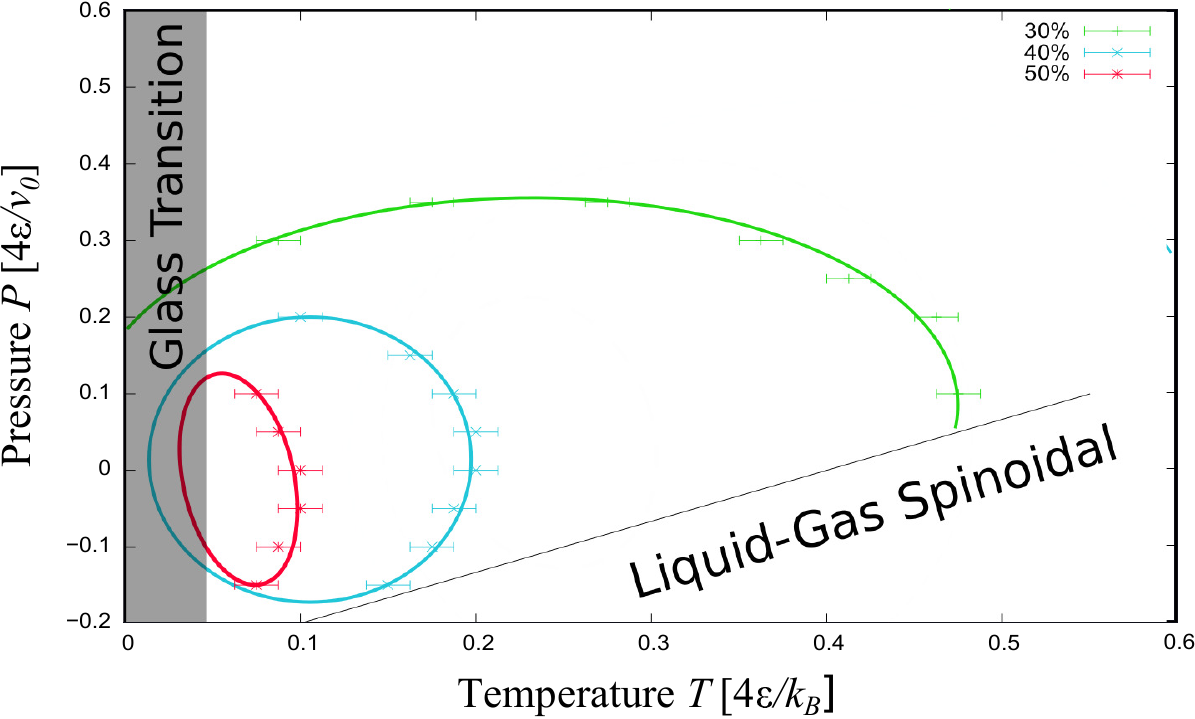}
\caption{{\bf Stability regions for the hydrophobic homopolymer in the hydrophobic slit pore without top-down symmetry}. Green, blue, and red symbols with error bars mark the state points where the chain has, on average, $N_{\rm CP}>$ 30\%, 40\%, and 50\%  $N_{\rm CP}^{\rm max}$, respectively. 
Elliptic lines are a guide for the eyes. 
The black line marks the liquid-to-gas spinodal for the confined water solution.
The grey region indicates the glassy state points at $k_BT/4\epsilon \lesssim 0.05$.}
\label{SR}
\end{figure}

These results are summarized in the $T$--$P$ thermodynamic plane as SRs (Fig.\ref{SR}). We find that the SRs at 30\%, 40\%, and 50\% $N_{\rm CP}^{\rm max}$ are concentric as expected \cite{Bianco:2015aa}.
The three SRs display a reentrant behavior in $T$ at different $P$, while the SR at 30\% also shows a reentrant behavior in $P$ at different $T$.
Each SR line can be adjusted to curves with different degrees of ellipticity, as expected by general arguments \cite{hawley1971}.
All the curves intersect the limiting temperature $k_BT/4\epsilon \lesssim 0.05$ below which we cannot equilibrate the system within our statistics (the grey region in Fig.\ref{SR}).
Moreover, the SR for 30\% intersects the liquid-to-gas spinodal line for our confined water solution. This line is marked by a significant volume increase of the entire system (not shown) and by discontinuities in $N_{\rm CP}$ for the four lowest pressures, reaching values typical of a random coil as at high $P$ (Fig.\ref{CP}). 

\begin{figure}
\centering
\includegraphics[width=\columnwidth]{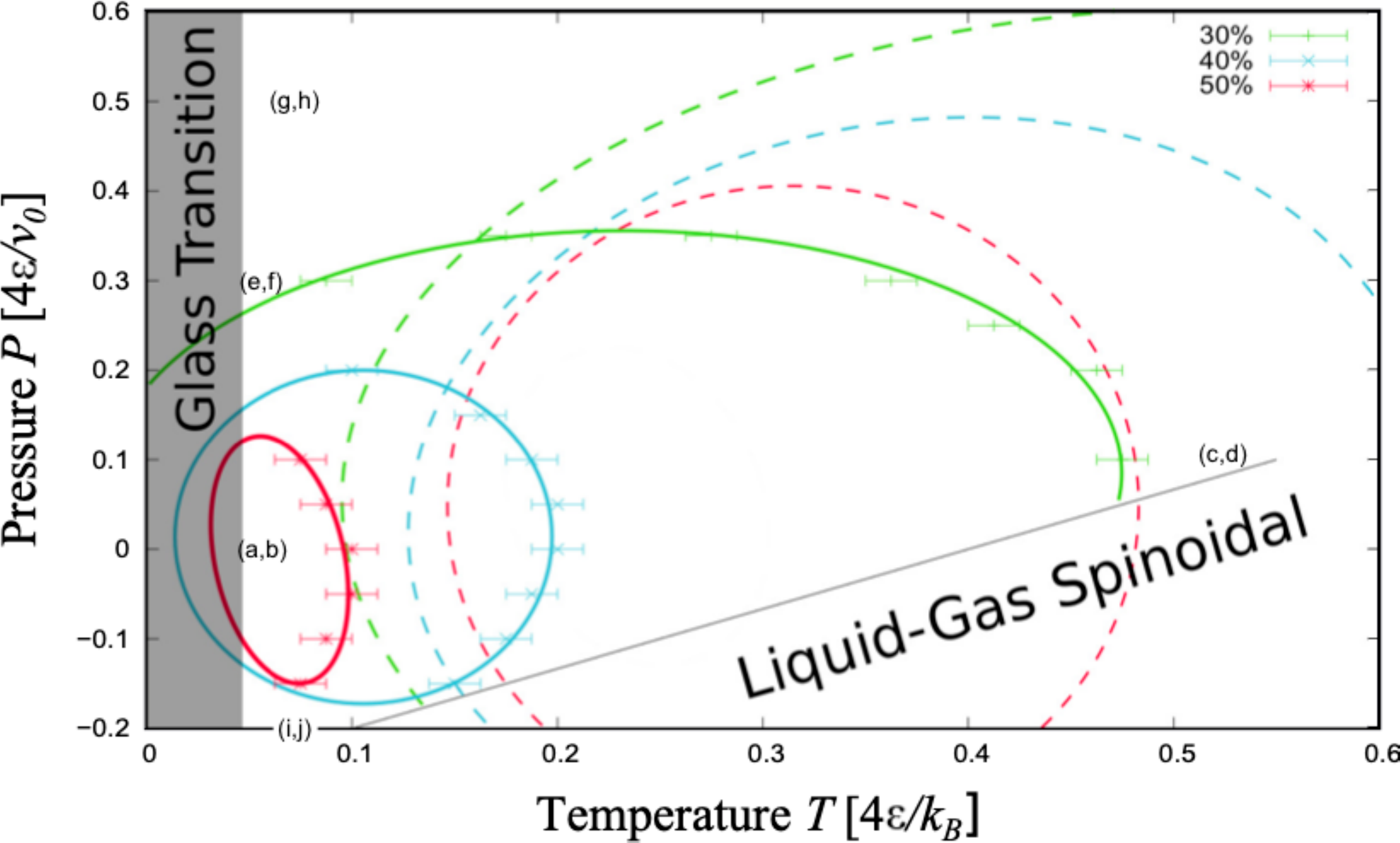}
\caption{{\bf The hydrophobic confinement without top-down symmetry destabilizes the globular conformations compared to the bulk case and affects the water liquid-to-gas spinodal.}
The latter (continuous black line) is shifted, at constant $P$, to lower $T$ by $\simeq 0.5 k_BT/4\epsilon$ relative to the bulk case (not shown here because out of scale).
The 30\%, 40\%, and 50\% SRs for the chain (continuous lines as in Fig~\ref{SR}) are displaced to lower ($T$, $P$) and are smaller compared to those for the free case (dashed lines with the same color code as the continuous). 
The grey area is as in Fig~\ref{SR}.
The labels (a,b), (c,d), etc., refer to the state points discussed in Fig.\ref{examples}.
All the lines are guides for the eyes.
The dashed lines are adapted with permission from Ref.~\cite{Bianco:2015aa} copyrighted (2015) by the American Physical Society \url{https://doi.org/10.1103/PhysRevLett.115.108101}.
}
\label{comparison}
\end{figure}

\begin{figure}
\subfloat[]{\includegraphics[width = 1.116in]{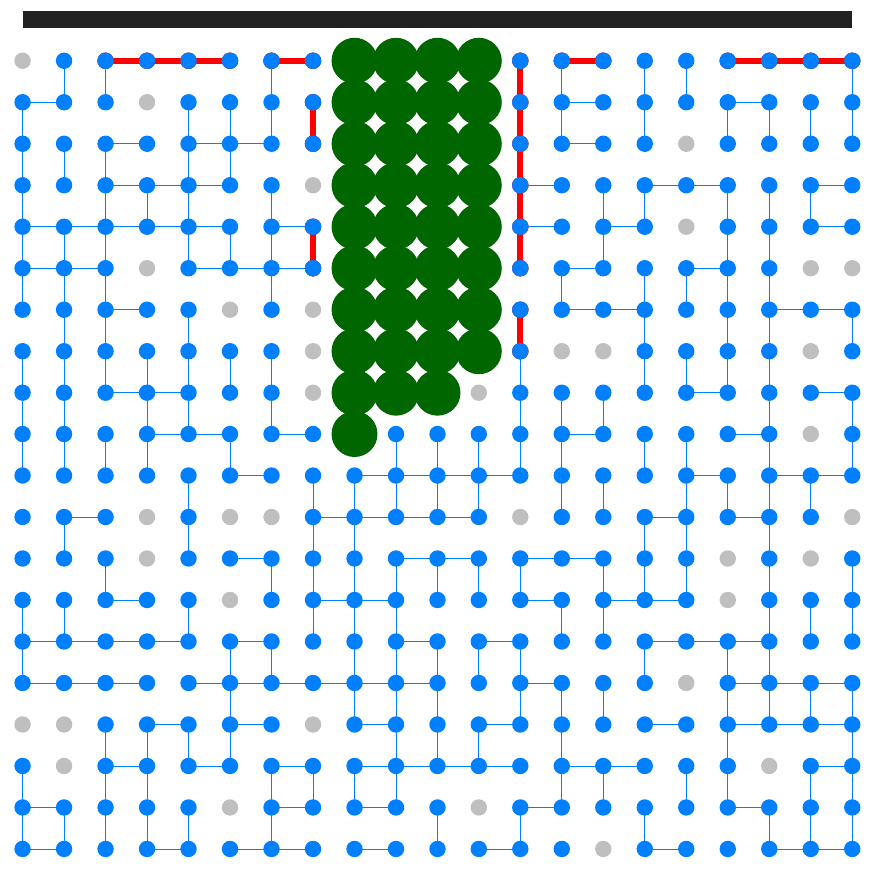}}
\subfloat[]{\includegraphics[width = 1.2753in]{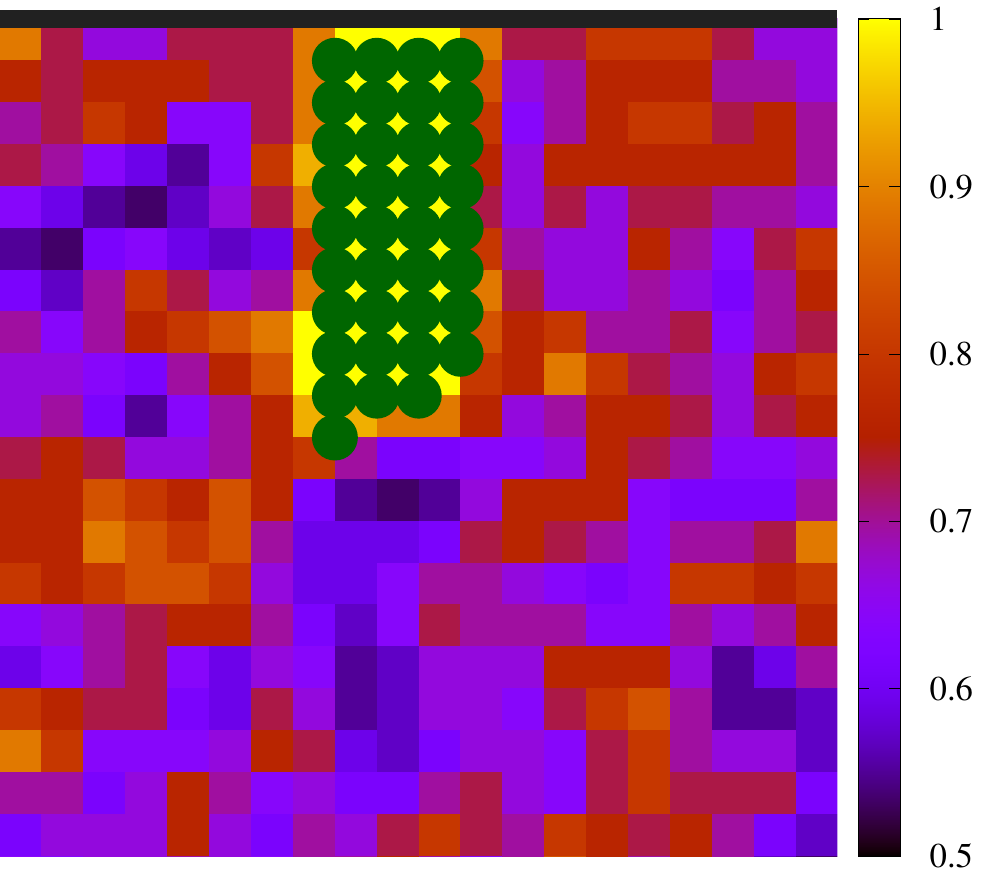}}\\
\subfloat[]{\includegraphics[width = 1.116in]{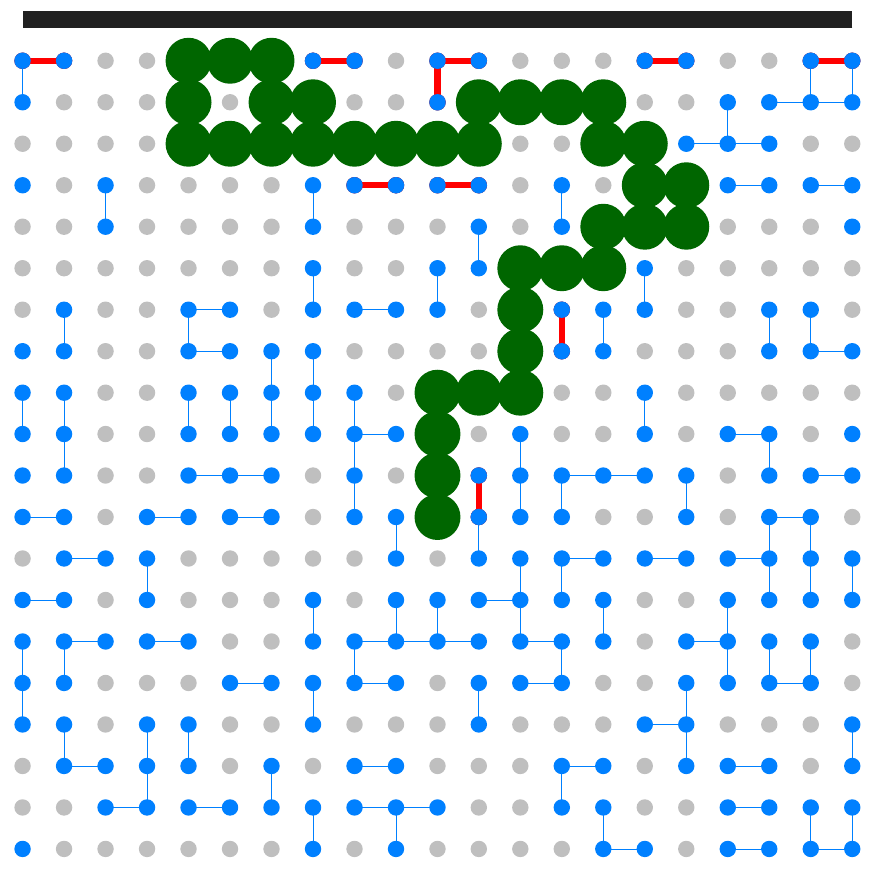}}
\subfloat[]{\includegraphics[width = 1.2753in]{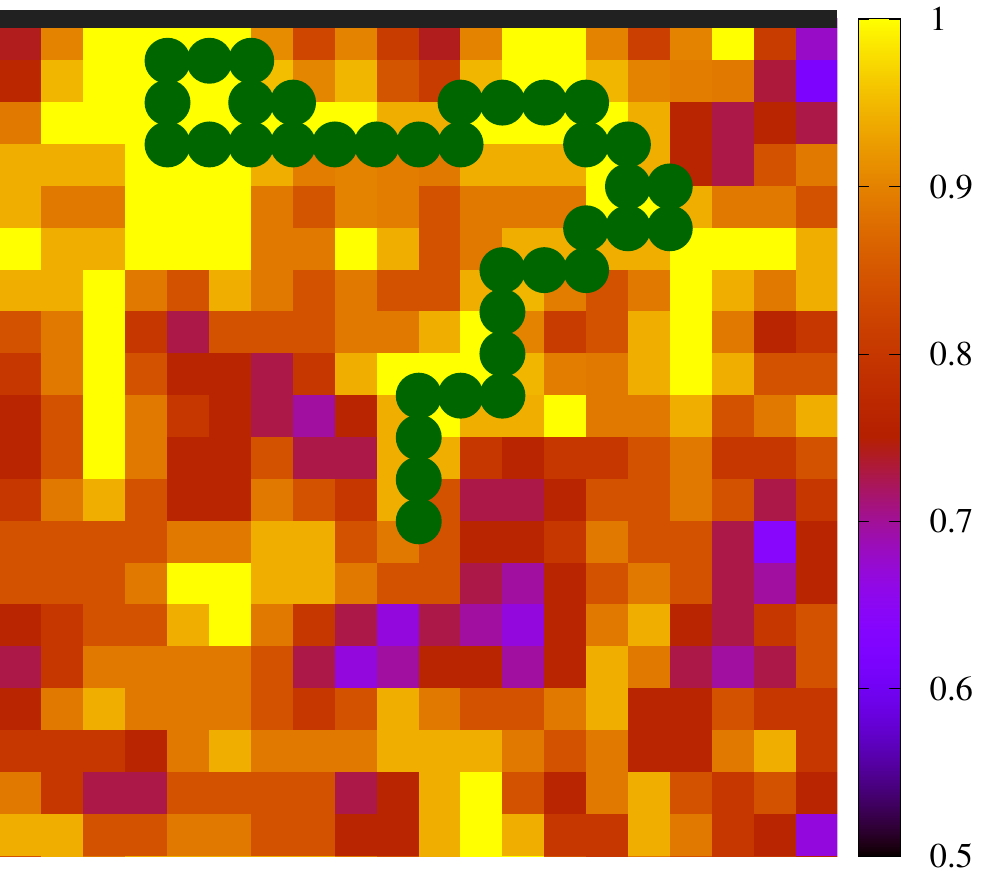}}\\
\subfloat[]{\includegraphics[width = 1.116in]{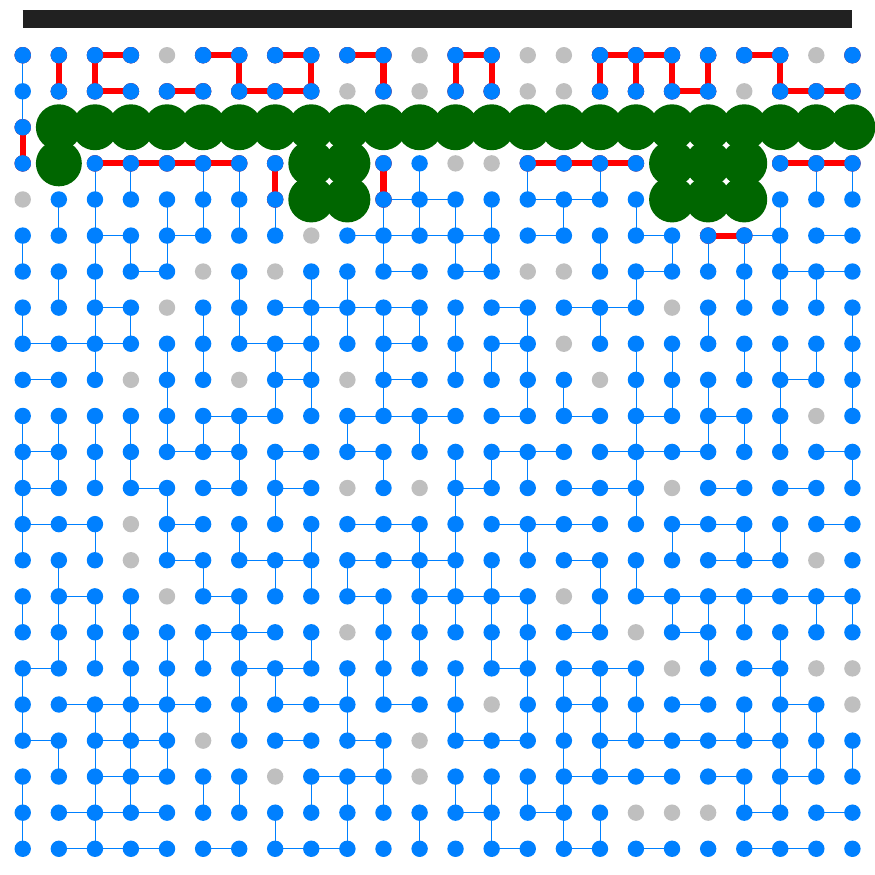}}
\subfloat[]{\includegraphics[width = 1.255in]{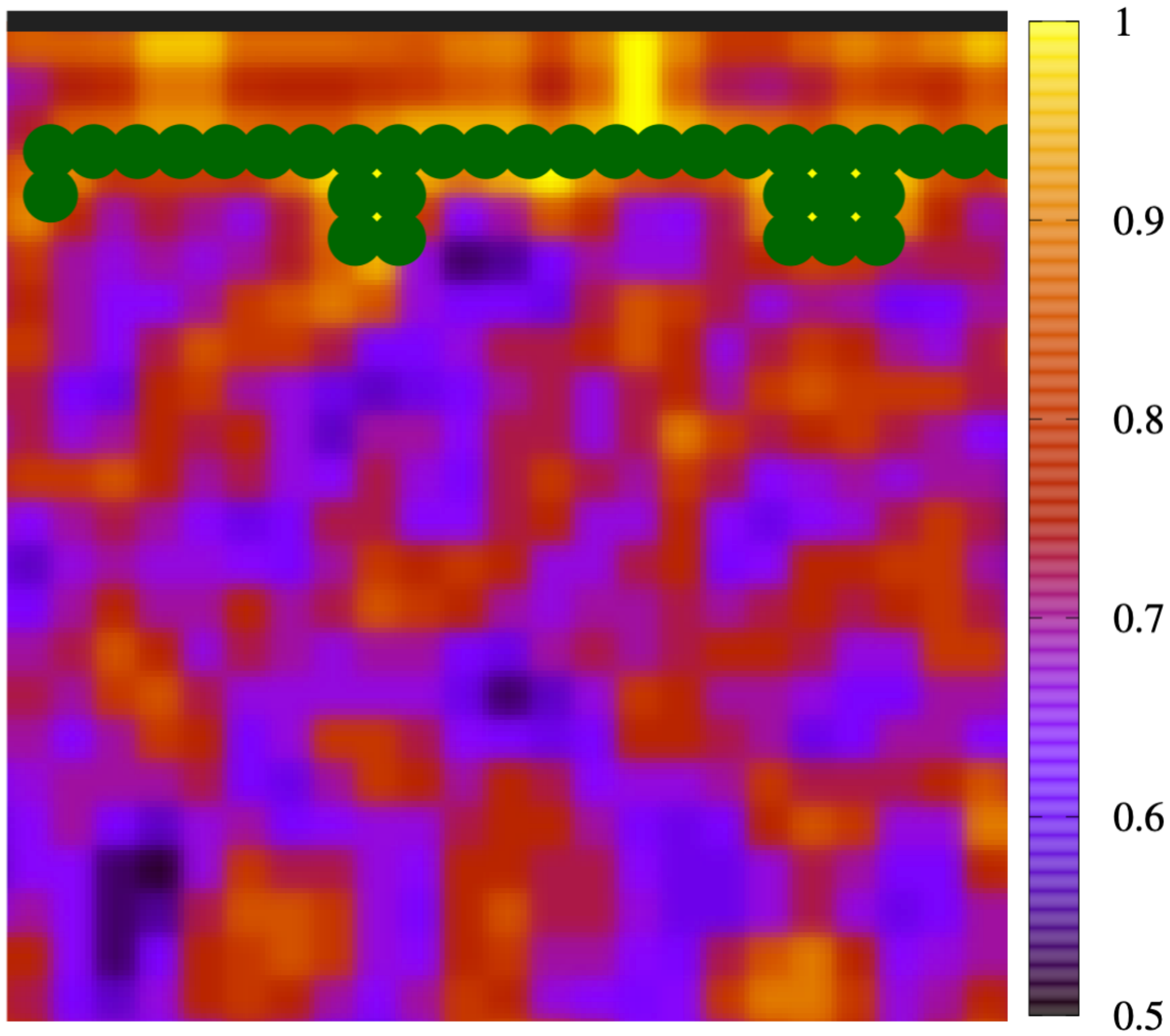}}\\
\subfloat[]{\includegraphics[width = 1.116in]{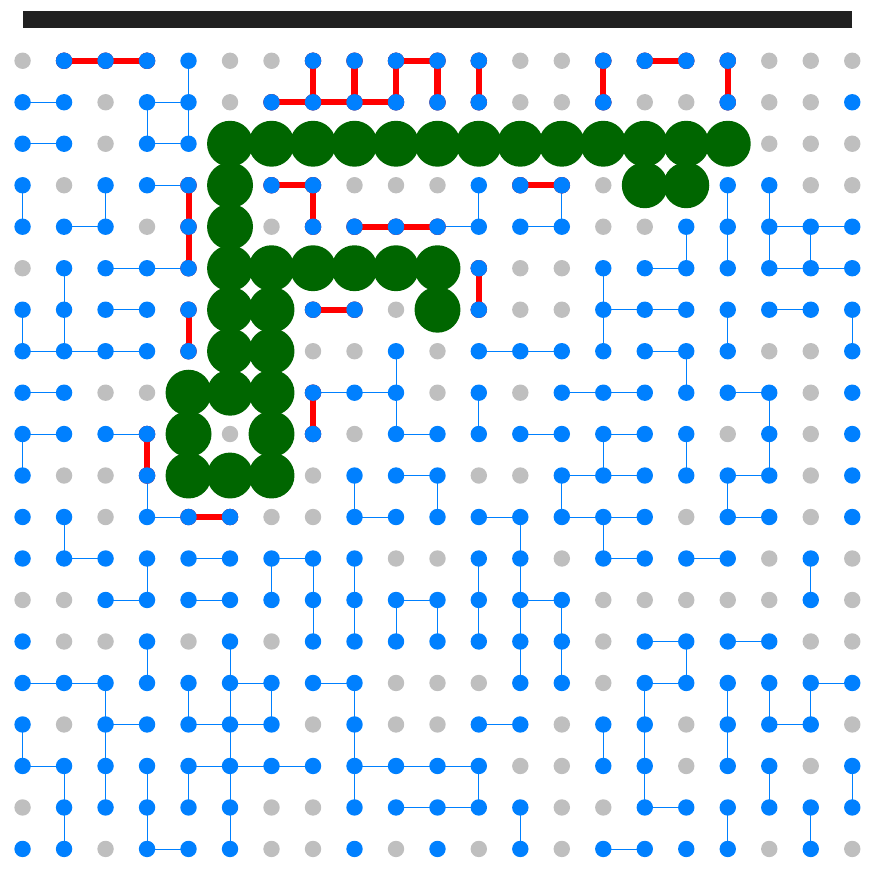}}
\subfloat[]{\includegraphics[width = 1.2753in]{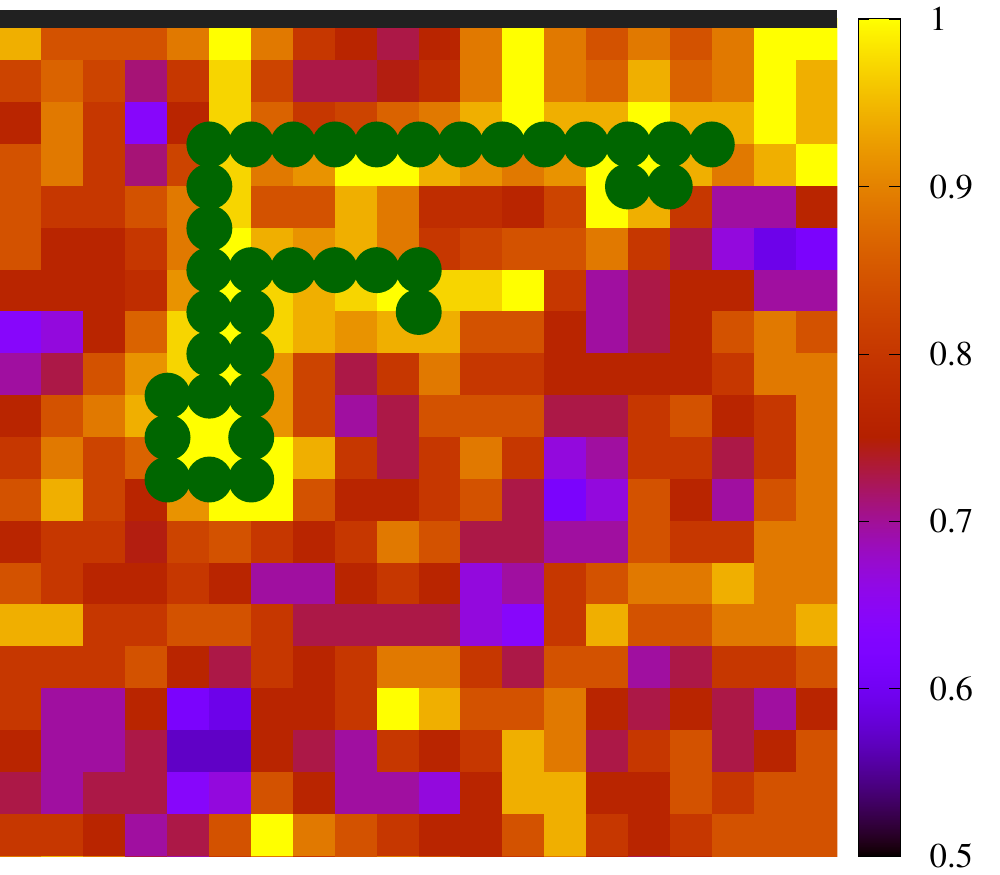}}\\
\subfloat[]{\includegraphics[width = 1.116in]{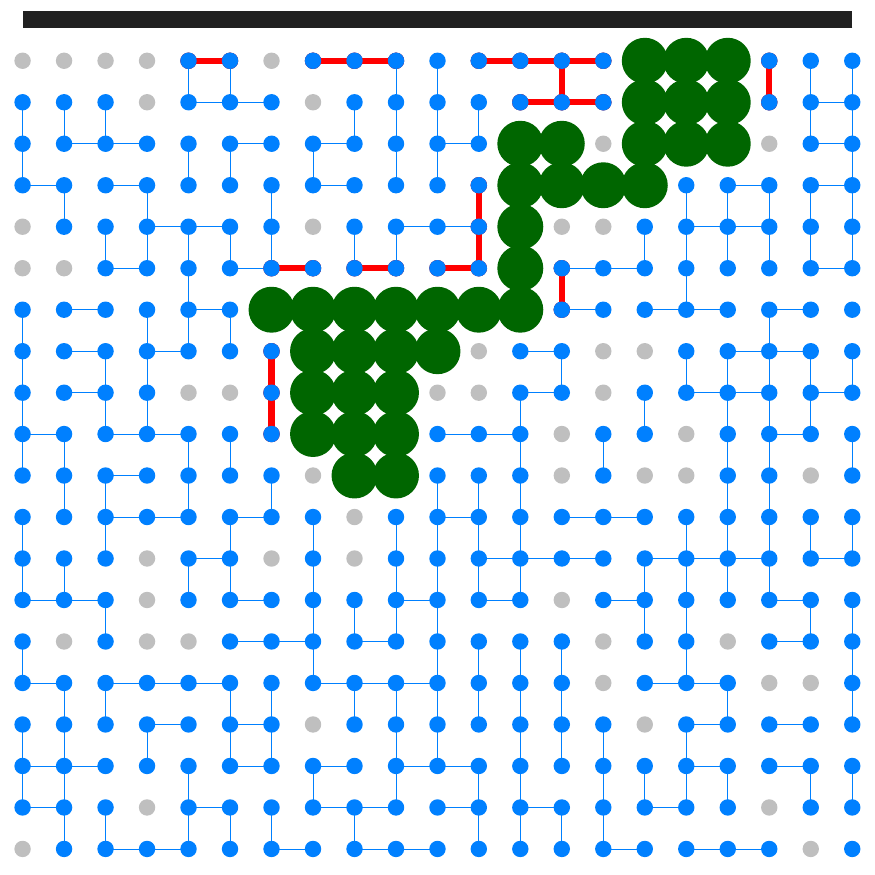}}
\subfloat[]{\includegraphics[width = 1.2753in]{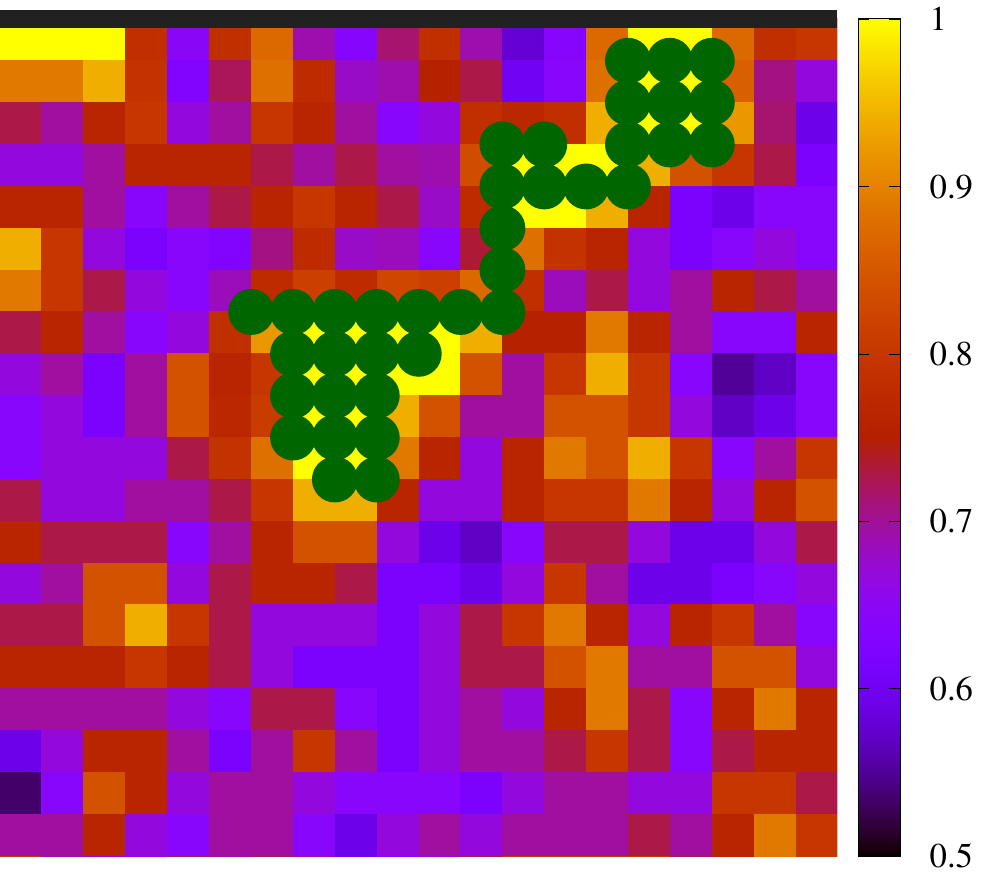}}\\
\caption{{\bf Coil-globule transition and adsorption without top-down symmetry at different state points}.  
Left panels:
HB network of bulk (blue) and hydration (red) water.
Right panels: Color-coded water density field from low (dark blue) to high  (yellow).  
The regions with no HBs (grey dots on the left panels) have a higher density (yellow regions on the right panels).
The thermodynamic state points $(k_BT/4\epsilon, Pv_0/4\epsilon)$ of the panels are reported in the phase diagram in Fig.\ref{comparison}:
{\bf (a), (b)} (0.050, 0.0);
{\bf (c), (d)} (0.525, 0.1);
{\bf (e), (f)} (0.050, 0.3);
{\bf (g), (h)} (0.075, 0.5);
{\bf (i), (j)} (0.075, -0.2).
The proteins-like chain and the interface are represented similarly as in Fig.\ref{system}.
}
\label{examples}
\end{figure}

\subsection{Comparison with the transition without the slit pore}

The hydrophobic confinement without top-down symmetry affects both the water and the protein. It changes the limit of stability (spinodal) of the liquid water 
compared to the gas phase (Fig.\ref{comparison}). At fixed pressure, we find the spinodal at lower $T$ than the free chain case \cite{Bianco:2015aa}. Overall the new spinodal is parallel to the former with a shift to lower $T$ of $\simeq 0.5 k_BT/4\epsilon$ at constant $P$. This effect is independent of breaking the top-down symmetry (Fig.S1).

This result is a consequence of the interaction of the liquid with the slit pore. Confinement generally affects the properties of liquids, particularly water \cite{Calero2016, MCF2017, Martelli:2017aa, Martelli:2020ab, Corti:2021uy, Leoni:2021aa}.
The effect of hydrophobic evaporation has been extensively studied for confined water, e.g., in Ref.~\cite{Remsing:2015aa} and references therein.
Near ambient conditions, water dewets the walls of a hydrophobic nano-pore and evaporates \cite{Sharma:2012aa}. 
Experiments show capillary evaporation at scales consistent with the size of our slit pores at lower $P$ and higher $T$ compared to ambient conditions \cite{capillary2014}.

Moreover, the presence of hydrophobic walls without top-down symmetry 
modifies the SRs compared to the free case (Fig.\ref{comparison}). 
We find two striking features. First,
all the regions marking 30\%, 40\% and 50\% of $N_{\rm CP}^{\rm max}$ for the confined chain occur at values of $T$ and $P$ that are lower than those for the free case \cite{Bianco:2015aa}.
Second, the confined protein-like polymer has a coil-globule transition in a $(T, P)$ range much smaller than the free case \cite{Bianco:2015aa}.
As a consequence of these changes, if a free chain comes into contact with 
the biased hydrophobic surface at a thermodynamic condition where it is in a globule state, e.g., $(Tk_B/4\epsilon, Pv_0/4\epsilon)$ = (0.4, 0.1), its $N_{\rm CP}$ would reduce from more than 50\% to 30\% of $N_{\rm CP}^{\rm max}$ (Fig.\ref{comparison}).

To check the effect of the bias, we repeat the calculations for a slit pore with top-down symmetry and find no differences for the confined water phase diagram, while we observe that the change in the SR compared to the bulk case is negligible (Fig.S1). Furthermore, to check the effect of the energy gain for the HB at the hydrophobic interface, we also decrease the $\Delta J^{(\phi)}/J$ parameters in the unbiased case (Table~S1). This change implies that the SR shifts to lower $T$ and is less accessible than the case with larger $\Delta J^{(\phi)}/J$. Consequently, the unfolding at low $T$ and low $P$ are not observed for the weak $\Delta J^{(\phi)}/J$.

Hence, our results suggest that facilitated adsorption, e.g., due to an attractive force or a drift toward the interface, significantly destabilizes the globular conformations of the polypeptide.
As a further confirmation of this observation, we find that the maximum number of CPs that the chain reaches in the biased hydrophobic slit pore is 55\% of $N_{\rm CP}^{\rm max}$, while it is more than 70\% for the free case \cite{Bianco:2015aa}.

\subsection{Interplay of adsorption and coil-globule transition}

The thermodynamic state-point affects not only the coil-globule but also the adsorption-desorption transition of the hydrophobic homopolymer on the hydrophobic surface, showing an intriguing interplay between the two phenomena.

\subsubsection{Adsorption in the globule state}

At low $T$ and $P$,  e.g., at $(k_BT/4\epsilon, Pv_0/4\epsilon)=(0.050, 0.0)$, one expects that the most relevant term in the BF Gibbs free energy $G^{(BF)}$, Eq.(\ref{GBF}), is the interaction energy, ${\cal H}_{\rm TOT}$, 
while both $TS_{\rm TOT}$ and $PV_{\rm TOT}$ contributions are vanishing. 
Because ${\cal H}_{\rm TOT}$ is dominated by the $N_{\rm HB}$ term, the $G^{(BF)}$ minimum corresponds to a maximum in $N_{\rm HB}$. 
Hence,  the unstructured chain adsorbs onto the surface, allowing more water molecules to form bulk HBs. 

The water release induces, macroscopically, an effective hydrophobic attraction between the surface and the residues. 
As expected for the low relevance of the entropic and volumic terms in $G^{(BF)}$ under these conditions, the many HBs
organize in a highly ordered network with low $S_{\rm TOT}$ (Fig.\ref{examples} a) and large volume (Fig.\ref{examples} b),
independent of the presence of bias (Fig.S2 a, b). In particular, for the unbiased case, the protein adsorbs onto the hydrophobic interface when it is in its globule state (Fig.S3).

This result is consistent with experiments. For example, blood 
proteins adsorb and form a corona onto nanoparticles with hydrophobic patches \cite{Soddu2020}.
The common understanding is that the effect is maximum if the proteins flatten onto the nanomaterial \cite{Rabe:2011kx, Wheeler2021}. 
Yet, experiments show that at least a part of the proteins in the corona can retain their functional motifs to allow the receptors' recognition  \cite{Kelly:2015aa, Bertoli:2016aa, Lara:2017aa} especially {\it in vivo} \cite{Hadjidemetriou:2015aa}. In particular, the IDRs of structured proteins can be almost unaffected in their globular state when adsorbed onto a surface \cite{Desroches:2007aa}.

Our results offer a rationale for this surprising experimental result. Indeed, we observe that the adsorbed homopolymer often keeps a globule conformation at $T$ and $P$ within its SR, as shown in movies mov1.mp4 and mov1nobias.mp4 in Supporting Information (SI) for biased and no-biased slit-pore, respectively. 
This is because 
${\cal H}_{\rm TOT}$ is minimized when both $N_{\rm HB}$ and $N_{\rm HB}^{(\phi)}$, i.e., the number of HBs in bulk and within the hydration shell, respectively, are maximized. 
Hence, the chain adsorps onto the surface to maximize $N_{\rm HB}$ but leaves as much as possible of the hydrophobic interface exposed to water to maximize $N_{\rm HB}^{(\phi)}$
(Fig.s \ref{examples} a, and S2 a).

\subsubsection{Adsorption in the coil state at high $T$}

At higher $T$, approaching the liquid-gas spinodal, e.g., at $(k_BT/4\epsilon, Pv_0/4\epsilon)=(0.525, 0.1)$, the entropy dominates the Gibbs free energy, Eq.(\ref{GBF}), and the homopolymer loses its
globule conformation, increasing $S_{\rm TOT}$
(Fig.~\ref{examples} c, d for the biased case and Fig.S2 c, d for the unbiased). 
Most of the time, the chain is kept adsorped onto the biased surface without top-down symmetry (mov2.mp4 in SI), while it is free when the biased is absent (mov2nobias.mp4 in SI).

Hence, the thermodynamic state point controls
how much the adsorbed polypeptide collapses or coils. Furthermore, it is reasonable to suppose that other relevant control parameters are the biomolecule and interface hydrophobicities, although we do not vary them here. 

\subsubsection{Desorption in the coil state at low $T$}

The BF model shows that in bulk, at low enough $T$ and appropriate $P$, the coil-globule transition is reentrant \cite{Bianco:2015aa} (Fig.~\ref{comparison}) as seen in experiments, at relatively high pressures, in structured proteins \cite{Pastore2007, Adrover:2012aa, Sanfelice:2015aa, Yan:2018aa}, and unstructured polymers \cite{Kunugi2002329}. 
Furthermore, recent experiments show that the folded domains of fused in sarcoma (FUS), a protein with low-complexity IDRs, undergo cold denaturation, with implications for its mediation of LLPS \cite{Felix:2023aa}.

Here, we observe for the hydrophobic polymer under biased confinement the analogous of the cold denaturation at low $T$ and a high enough $P$ (Fig.~\ref{CP}), e.g., at $(k_BT/4\epsilon, Pv_0/4\epsilon)=(0.050, 0.3)$ (Fig.~\ref{comparison}).
This low-$T$ unfolding is energy-driven due to the contribution of the hydration HBs to the ${\cal H}_{\rm TOT}$ \cite{Bianco:2015aa} and is unaccessible if the 
energy gain of the HB at the hydrophobic interface is too small (Fig.S1).

We find that, at the reentrant transition, the chain often extends and desorbs from the biased hydrophobic interface (Fig.\ref{examples} e, f, and mov3.mp4 in SI). 
Intriguingly, the polymer flattens out, keeping a characteristic distance from the interface of two layers of water. As a consequence, it minimizes ${\cal H}_{\rm TOT}$ by
maximizing the $N_{\rm HB}^{(\phi)}$ at the two hydrophobic interfaces--the homopolymer and the wall--and the bulk $N_{\rm HB}$.

This observation is consistent with atomistic simulations showing that the bilayer is the most stable free-energy minimum for water confined in a hydrophobic slit pore \cite{calero2020}. 
Furthermore, this minimum is energy-driven by the water HBs that saturate to their maximum number per molecule \cite{Leoni:2021aa}. Therefore, the BF model captures the atomistic features of the energy-driven double-layer of water while showing the low-$T$ flattening of the polymer and its desorption from the interface. 

Interestingly, simulations of coarse-grained hydrophobic IDPs in implicit water with effective (water-mediated) $T$-dependent interactions display an upper critical solution temperature (UCST) and a lower critical solution temperature (LCST) \cite{Dignon:2019to}, as in experiments with designed IDPs \cite{Quiroz:2015aa}. Here, the BF model with the reentrant coil-globule transition for a hydrophobic polymer offers an ideal test for this phenomenology without introducing effective $T$-dependent interactions, being transferable and water-explicit.

\subsubsection{Desorption in the coil state under pressurization}

At large $P$, e.g., $(k_BT/4\epsilon, Pv_0/4\epsilon)=(0.075, 0.5)$, the Gibbs free energy, Eq.(\ref{GBF}), is dominated by the volume term. As discussed for the bulk case  \cite{Bianco:2015aa},  in agreement with the experiments for protein $P$-induced unfolding \cite{MeersmanFilip2006}, the large compressibility of the hydration water at hydrophobic interfaces allows the system to reduce the $V_{\rm TOT}$ under pressurization. Hence, the chain undergoes a density-driven transition from a globule to a coiled state, as shown by the high-density regions we find around the polymer under these thermodynamic conditions (Fig.\ref{examples} g, h and Fig.S2 e, f). Furthermore, the high $P$ induces a decrease in water HBs number \cite{FS2007}, diminishing the effective hydrophobic attraction between the surface and the polypeptide, leading to desorption even in the biased case (mov4.mp4 in SI). 

This finding calls for experiments on the protein corona formation and evolution onto nanoparticle and nanomaterials under pressure changes. While $T$ effects are known in the corona composition \cite{Mahmoudi:2022aa}, to our knowledge, no studies are available as a function of pressure.

\subsubsection{Adsorption in the coil state under tension}

Under tension, e.g., $(k_BT/4\epsilon, Pv_0/4\epsilon)=(0.075, -0.2)$,  we find that the chain unfolds but is still adsorbed onto the biased hydrophobic surface (Fig.\ref{examples} i, j, and mov5.mp4 in SI). 
From Eq.(\ref{GBF}), we observe that the Gibbs free energy in this thermodynamic regime is minimized by maximizing the volume. From the definition of $V_{\rm TOT}$, we note that this condition corresponds to maximizing both $N_{\rm HB}$ and $N_{\rm HB}^{(\phi)}$ at $P<0$. Therefore, the polymer loses its globule state, exposing the hydrophobic residues to hydration. 
However, the $P<0$ unfolding occurs only if the energy gain at the hydrophobic hydration is large enough. Indeed, for the unbiased case with small $\Delta J^{(\phi)}/J$ (Table~S1) the unfolding at negative $P$ is not accessible (Fig.S1).

Under tension, the degree of unfolding is moderate compared to the other cases (at low-$T$, high-$P$, or high-$T$) because a large stretch of the polypeptide would imply an increase of hydration water with large compressibility, inducing a decrease of $V_{\rm TOT}$. Consequently, the protein-like chain explores conformations that compromise between globular and unfolded regions. 

At the same time, the increase in the number of HBs implies a strengthening of the water-mediated hydrophobic attraction between the homopolymer and the surface and the consequent adsorption onto the wall. 
This effect is also evident in the unbiased case. The chain diffuses slowly but, once near the surface, adsorbs irreversibly within our simulation time (mov3nobias.p4 in SI for the protein at $(k_BT/4\epsilon, Pv_0/4\epsilon)=(0.15, -0.1)$ under confinement with top-down symmetry).

Therefore, the unfolding and adsorption of the hydrophobic homopolymer are enthalpy driven.
These observations are possibly relevant in the force-induced protein unfolding and the LLPS under mechanical stress. Cells are permanently exposed to stress resulting from mechanical forces such as, e.g., the tension generated inside adherent and migrating cells,  sufficient to unfold cytoskeleton proteins \cite{Hohfeld:2021aa}. Under these tensile conditions, the unfolded proteins can aggregate \cite{Bianco:2020aa}, interfering with essential cellular processes and causing severe pathologies--such as neurodegenerative diseases and dementia \cite{Hipp:2019aa}--for which mechanopharmacology is emerging as a possible control strategy \cite{C7CS00820A}.

\section{Conclusions}

We study a coarse-grained hydrophobic homopolymer chain in a hydrophobic slit pore as a minimal model of an IDP near an interface in a spirit similar to that of Ref. \cite{Statt:2020vu}, choosing the entirely hydrophobic sequence to emphasize the effective hydrophobic interaction with the surface.
We use the BF model in explicit water and perform Monte Carlo free energy calculations under different thermodynamic conditions in confinement with and without top-down symmetry, the latter case mimicking a drift or a weak force pushing the protein toward the interface without limiting its lateral diffusion.
Our results reveal that the biased hydrophobic walls drastically affect the coil-globule transition of the polymer, reducing its stability region and shifting it to lower $T$ and $P$.

We find an intriguing interplay between the surface adsorption-desorption and the coil-globule transition. 
A protein unfolds partially when it approaches the surface \cite{Vilanova2016, Park:2021ve}. However, we find that 
the homopolymer can adsorb onto the hydrophobic interface, keeping, at least in part, a globule conformation consistent with recent protein {\it in vitro} \cite{Desroches:2007aa} and {\it in vivo} experiments \cite{Hadjidemetriou:2015aa}. 
 
At high $T$, the entropy drives the unfolding of the chain, but not necessarily its desorption when the bias is present. This result is of particular interest in developing 
strategies based on hyperthermia with protein-functionalized magnetic nanoparticles brought, under the action of forces resulting from external magnetic fields, to high $T$ for local treatments of, e.g., cancer cells \cite{Tay:2018aa}.

A similar result is also valid when the chain is under (mechanical) tension. It unfolds but does not necessarily desorbs from the surface. Under these circumstances, the polymer has a less extended conformation, where elongated regions intercalate small globules, keeping their adhesion to the interface. Understanding this mechanism could be crucial to treat diseases involving junctions \cite{Getsios:2015aa} as, e.g., cardiac disorders \cite{Noorman:2009aa},  where mechanical forces trigger the loss of tertiary and secondary structural elements within anchoring proteins \cite{Hohfeld:2021aa}.

Under high-pressure and, possibly, low-temperature stresses, chains lose their globular state and desorb from the hydrophobic interface, typically separated by a water bilayer, driven by water's density at high $P$ and water's energy at low $T$. 
The energy-driven low-$T$ behavior is consistent with 
atomistic simulations showing that the bilayer is the most stable free-energy minimum for hydrophobically confined water  \cite{calero2020}.
Also, it offers an ideal test with a transferable and water-explicit
molecular model for recent IDPs {\it in vitro} experiments \cite{Quiroz:2015aa} and 
coarse-grained IDPs implicit-water simulations, with effective $T$-dependent interactions, displaying  LLPS with UCST and LCST \cite{Dignon:2019to}.
At the same time, our predictions call for new experiments on the protein corona evolution on nanomaterials under pressurization.

\section*{Supporting Information}

Additional simulations details, including 
{\bf 1)} a table with the model’s parameters with or without top-down symmetry (Table S1),
{\bf 2)} data for the unbiased confinement
(Figures S1, S2, S3),
and
{\bf 3)} movies for the biased confinement 
(MP4 movies mov1.mp4, mov2.mp4, mov3.mp4, mov4.mp4, mov5.mp4),
and the unbiased confinement 
(MP4 movies mov1nobias.mp4, mov2nobias.mp4, mov3nobias.mp4).

\section*{Acknowledgement}

We want to acknowledge Prof. Pablo G. Debenedetti
for the fruitful discussions about water and biological systems over the years. %
G.F. acknowledges support from the Spanish grants PGC2018-099277-B-C22 and PID2021-124297NB-C31, funded by MCIN/AEI/ 10.13039/ 501100011033 and ``ERDF A way of making Europe", and the Visitor Program of the Max Planck Institute for The Physics of Complex Systems for supporting a six-month visit started on November 2022.

\medskip

\bibliography{bibliografia}

\end{document}